\documentclass{article}
\usepackage{spconf,amsmath,graphicx,hyperref}
\usepackage{multirow}
\usepackage{graphicx}

\usepackage{amsmath,amssymb} 
\usepackage{dsfont}          
\usepackage{float} 
\usepackage{subcaption}
\usepackage{booktabs}
\usepackage{siunitx}
\usepackage{dsfont}
\newcommand{\ind}[1]{\mathds{1}\!\left[\,#1\,\right]}

\usepackage{hyperref}
\hypersetup{hidelinks}

\title{R\textsuperscript{3}-REC: Reasoning-Driven Recommendation via Retrieval-Augmented LLMs over Multi-Granular Interest Signals}
\name{Yuchen Miao\qquad Mingxuan Cui\qquad Yitong Zhu\qquad Yu Wang\qquad
Siyang Xu}

\address{Sydney Smart Technology College, Northeastern University, China}

\begin{document}
\ninept
\maketitle
\begin{abstract}
This paper addresses two persistent challenges in sequential recommendation: (i) evidence insufficiency—cold-start sparsity together with noisy, length-varying item texts; and (ii) opaque modeling of dynamic, multi-faceted intents across long/short horizons. We propose \textsc{R\textsuperscript{3}-REC} (\textbf{R}easoning–\textbf{R}etrieval–\textbf{R}ecommendation), a prompt-centric, retrieval-augmented framework that unifies Multi-level User Intent Reasoning, Item Semantic Extraction, Long-Short Interest Polarity Mining, Similar User Collaborative Enhancement, and Reasoning-based Interest Matching and Scoring. Across \emph{ML-1M}, \emph{Games}, and \emph{Bundle}, \textsc{R\textsuperscript{3}-REC} consistently surpasses strong neural and LLM baselines, yielding improvements up to {+}10.2\% (HR@1) and {+}6.4\% (HR@5) with manageable end-to-end latency. Ablations corroborate complementary gains of all modules.
\end{abstract}

\section{Introduction and Related Work}
\label{sec:intro_related}

Sequential recommenders rank top-$k$ items from recent interactions for large platforms~\cite{Quadrana2018SARS}.
Yet two issues remain stubborn: (i) evidence insufficiency—cold-start sparsity together with noisy, length-varying item texts; and (ii) opaque modeling of dynamic, multi-faceted intents across long/short horizons.

\textbf{Related work.} 
We categorize prior art by modeling principle. 
Latent sequential encoders—RNN/Transformer pipelines (GRU4Rec, SASRec, BERT4Rec) and graph variants (SR-GNN, GCE-GNN)—scale but hide semantics and struggle under sparsity~\cite{Hidasi2016GRU4Rec,Kang2018SASRec,Sun2019BERT4Rec,Wu2019SRGNN,Wang2020GCEGNN}, prompting recent diffusion-based methods to address multimodal feature denoising~\cite{lu2025dmmd4sr,cui2025diffusion,cui2025multi}. 
Intent/multiinterest models (NARM, STAMP, MCPRN, HIDE, Atten-Mixer, MIND, ComiRec) disentangle factors to lift diversity, yet often fix intent cardinality or underuse textual cues~\cite{Li2017NARM,Liu2018STAMP,Wang2019MCPRN,Chai2022HIDE,Zhang2023AttenMixer,Li2019MIND,Cen2020ComiRec}. 
LLM based recommenders (NIR, PO4ISR, RecMind) add language priors via prompting or interactive generation to improve coherency~\cite{Hou2023NIR,Sun2024PO4ISR,Wang2024RecMind,xie2025chat,liu2025coherency}, while strong non-LLM regularization (UniRec) remains competitive~\cite{UniRec2024}. 
No strand jointly delivers fine-grained intent reasoning, robust text denoising, cold start resilience, and end-to-end explainability.

\textbf{Motivation and Our Approach.}
Embedding-only pipelines entangle evidence without rationale; multi-intent models fix capsule counts, limiting intent elasticity; graph encoders seldom refine noisy item texts; prompt-only LLMs reason over coarse user summaries. Guided by deliberative LLM recommendation \cite{Fang2025Reason4Rec}, retrieval-augmented sequential models \cite{Zhao2024RaSeRec}, and unified long/short-term preference modeling with RNNs \cite{Devooght2017LSTMRNN}, we design a prompt-centric, reasoning-augmented framework that separates intent inference, text refinement, and evidence aggregation before scoring: (a) multi-level intent reasoning with explicit polarity across long/short horizons; (b) item-text refinement with salient keyword extraction; (c) RAG-style similar-user retrieval to mitigate sparsity; and (d) reasoning-based interest–item matching that outputs faithful rationales under a training-light, modular pipeline.

\begin{figure}[t]
\centering
\includegraphics[width=0.7\columnwidth]{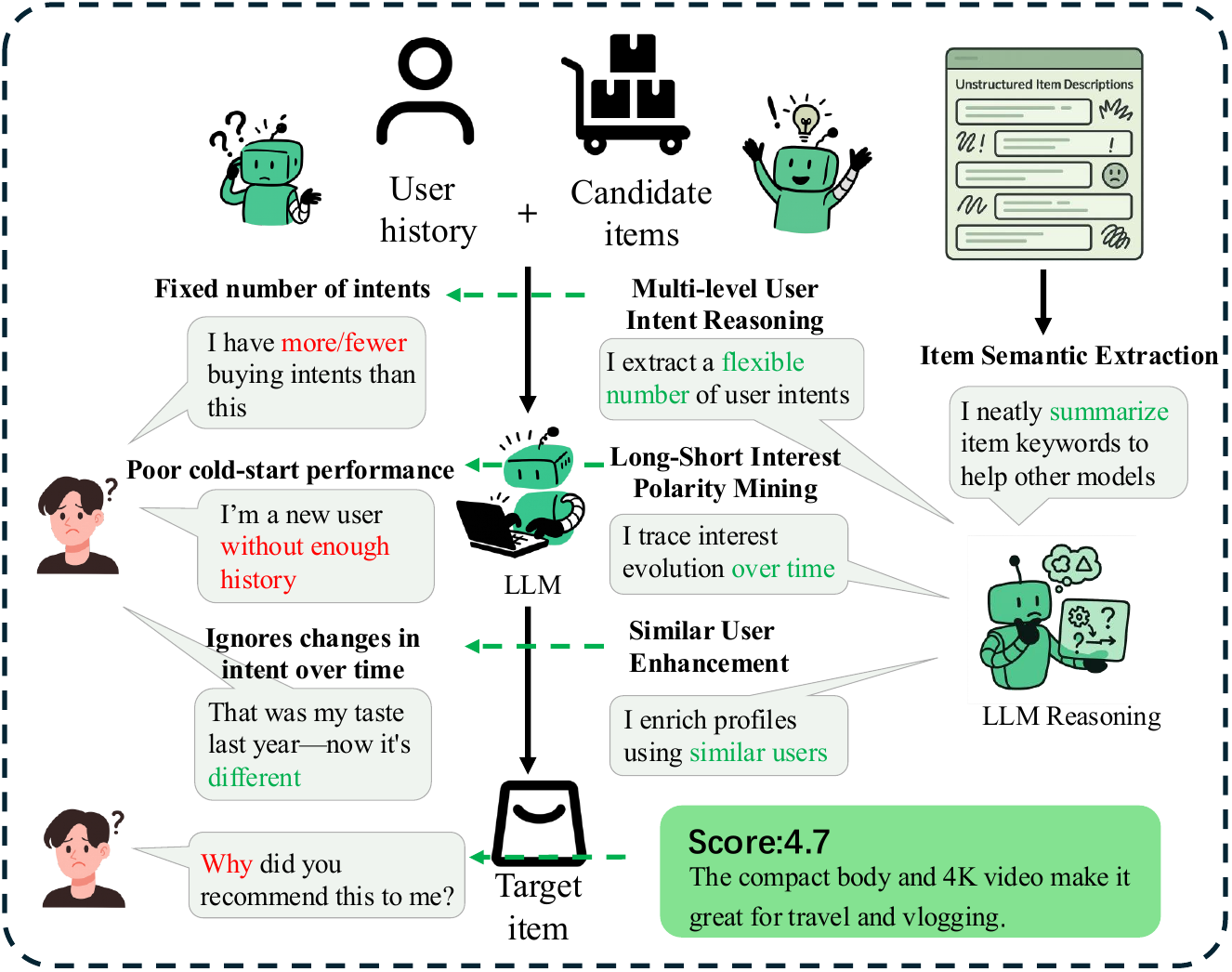}
\caption{Challenges and solution: (i) evidence insufficiency (cold-start + noisy item texts), (ii) opaque modeling of long/short-horizon intents.}
\label{fig:f1}
\end{figure}

\textbf{Contributions.}
(i) A reasoning driven paradigm that models elastic, multi-granular intents with transparent rationales.
(ii) A multi-route item-text refinement module that distills noisy metadata into compact, alignable semantics.
(iii) A similar-user collaborative enhancement stage that strengthens cold start and narrow history personalization without heavy fine-tuning.
(iv) Extensive experiments show consistent gains over neural and LLM baselines, with ablations confirming module complementarity and end-to-end latency within practical limits.

\section{METHODOLOGY}
\label{sec:method}

\begin{figure*}[!t]
    \centering
    \includegraphics[width=0.9\textwidth]{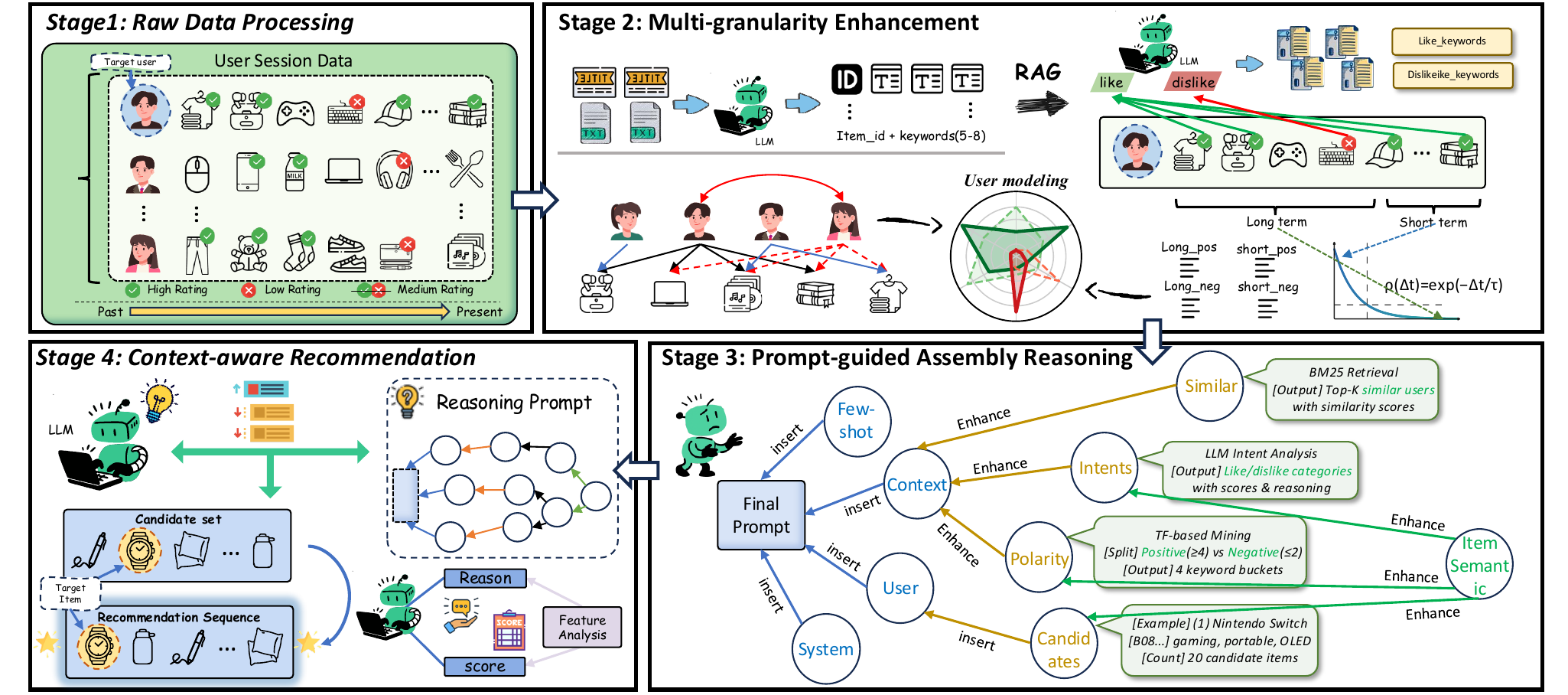}
    \caption{Methodology Overview of $R^3$-REC. The pipeline consists of four stages: 
    (1) \textbf{Raw Data Processing} handles user session streams; 
    (2) \textbf{Multi-granularity Enhancement} distills item semantics via LLMs and models hierarchical user interests (including intent reasoning and long-short term polarity mining); 
    (3) \textbf{Prompt-guided Assembly Reasoning} constructs a comprehensive prompt context by inserting retrieved similar users, inferred intents, and candidates; 
    (4) \textbf{Context-aware Recommendation} employs an LLM to perform feature analysis, generating both ranking scores and interpretative reasoning.}
    \label{fig:overview}
\end{figure*}

We propose $R^3$-REC, a reasoning-driven framework designed to bridge the gap between sparse sequential signals and the rich reasoning capabilities of Large Language Models (LLMs). As illustrated in Fig.~\ref{fig:overview}, our pipeline transforms raw interaction logs into a structured, retrieval-augmented context through four integrated stages: (1) extracting hierarchical user intents; (2) distilling item semantics; (3) retrieving collaborative evidence; and (4) performing reasoning-based scoring.

\subsection{Hierarchical Intent and Temporal Polarity Modeling}
\label{ssec:intent-polarity}
A core limitation of standard sequential models is their tendency to flatten user history into a single vector, obscuring the distinction between stable high-level interests (e.g., "RPG Games") and dynamic fine-grained preferences (e.g., "Cyberpunk Art Style"). To address this, we explicitly model user state at two granularities.

\subsubsection{Multi-level Recent Intent}
First, we capture the user's macro-level preference by analyzing their interaction distribution over item categories. However, simple frequency counts are susceptible to noise from impulsive clicks. To mitigate this, we propose a \emph{contrastive weighting mechanism} that penalizes categories with mixed feedback. For a user $u$ within a recent window $\mathcal{E}^{\mathrm{rec}}_u$, we compute the signed evidence $g_k^{(\pm)}$ for category $k$:
\begin{align}
g_k^{(\pm)} &= \sum_{e_j\in\mathcal{E}^{\mathrm{rec}}_u} \ind{i_j\in k}\,\rho(t_T - t_j)\,\ind{y_j=\pm 1}, \label{eq:gk} \\
\pi_k &= \frac{\exp\!\big(\kappa(g_k^{(+)}-g_k^{(-)})\big)}{\sum_{k'\in\mathcal{K}_u} \exp\!\big(\kappa(g_{k'}^{(+)}-g_{k'}^{(-)})\big)}, \label{eq:intent-weights}
\end{align}
where $\rho(\cdot)$ is a time-decay kernel ensuring recency, and $\kappa$ controls the sharpness of the distribution. By using the difference between positive and negative evidence $(g_k^{(+)}-g_k^{(-)})$, $\pi_k$ effectively suppresses categories where the user shows ambivalent behavior.
To ground these discrete weights in the dense representation space, we aggregate the pre-trained embeddings of categories $\mathbf{e}_k$ to form the \textbf{User Intent Vector}:
\begin{equation}
    \mathbf{z}_u = \sum_{k \in \mathcal{K}_u} \pi_k \mathbf{e}_k.
    \label{eq:zu_def}
\end{equation}
This vector $\mathbf{z}_u$ serves as a stable anchor representing "what" the user likes fundamentally.

\subsubsection{Long-Short Interest Polarity}
While categories define the domain, specific attributes drive the decision. We capture these micro-preferences by mining keywords from item metadata across two time horizons: short-term ($\mathcal{E}^{S}_u$, default 1 month) and long-term ($\mathcal{E}^{L}_u$, default 12 months).
For each horizon $\ell \in \{S, L\}$, we calculate a time-aware TF-IDF score $\omega^{(\ell)}(w)$ for each keyword $w$. Unlike standard TF-IDF, our formulation incorporates the temporal kernel $\rho(\cdot)$ to down-weight stale keywords. We then compute the \textbf{Polarity Vector} $\mathbf{q}^{(\ell)}_u$ by contrasting the centroids of positive and negative keywords:
\begin{equation}
\mathbf{p}^{(\ell)}_{\pm}=\frac{\sum_{w}\omega^{(\ell)}(w)\mathbf{e}_w}{\sum_{w}\omega^{(\ell)}(w)}, \quad
\mathbf{q}^{(\ell)}_u=\mathbf{p}^{(\ell)}_{+}-\mathbf{p}^{(\ell)}_{-}.
\label{eq:polarity}
\end{equation}
This vector explicitly encodes the directional shift in user taste (e.g., from "Casual" to "Hardcore").

Finally, we fuse these components into a unified query state $\mathbf{h}_u = \mathbf{W}_h [\mathbf{z}_u; \mathbf{q}^{(S)}_u; \mathbf{q}^{(L)}_u] + \mathbf{b}_h$.
\emph{Implementation Note:} Since the downstream LLM backbone is accessed via API and cannot be back-propagated, the projection matrix $\mathbf{W}_h$ acts as a lightweight adapter. To ensure $\mathbf{h}_u$ is semantically meaningful for the subsequent retrieval phase, we optimize $\mathbf{W}_h$ in a distinct \textbf{pre-training stage} using a soft alignment loss $\mathcal{L}_{\text{align}}$, which maximizes the cosine similarity between the categorical intent $\mathbf{z}_u$ and the attribute-based polarity $\mathbf{q}^{(\ell)}_u$.

\subsection{Item Semantic Extraction}
\label{ssec:item-semantic}
Incorporating full item descriptions (titles, attributes, abstracts) into the LLM prompt is often impractical due to token limits and the token limits and noisy descriptions often dilute salient signals in prompts. Therefore, we design a module to distill items into a compact set of $5\text{--}8$ high-value keyphrases.

We adopt a \emph{Budgeted Coverage Maximization} strategy that blends statistical rigor with semantic understanding. First, we generate a broad candidate pool $\mathcal{U}_i = \mathcal{C}_i \cup \mathcal{L}_i$, where $\mathcal{C}_i$ represents noun phrases extracted via PMI-based chunking, and $\mathcal{L}_i$ contains abstractive tags generated by an instruction-tuned LLM.
To select the optimal subset $S_i$, we avoid simple top-$k$ selection which often yields redundant phrases. Instead, we formulate this as a Facility Location problem:
\begin{equation}
\max_{S_i \subseteq \mathcal{U}_i} \left( \sum_{c \in S_i}\psi(c) + \lambda \sum_{q \in \mathcal{U}_i} \max_{c \in S_i}\cos(\mathbf{v}_q, \mathbf{v}_c) \right).
\label{eq:fl}
\end{equation}
Here, the first term maximizes the relevance score $\psi(c)$ (combining TF-IDF and embedding similarity), while the second term maximizes the semantic coverage of the selected set $S_i$ over the entire pool $\mathcal{U}_i$. This greedy optimization ensures that the final keyphrases provide a comprehensive yet non-redundant summary of the item, serving as reliable evidence for reasoning.

\subsection{Similar User Collaborative Enhancement}
\label{ssec:sim-user}
Sequential models often struggle with cold-start users or sparse histories due to a lack of sufficient signal. Collaborative filtering addresses this by leveraging the behavior of similar users. However, implicit latent vectors from traditional CF are not human-readable. In $R^3$-REC, we retrieve \emph{explicit textual summaries} from similar users to provide the LLM with interpretable references.

\textbf{Memory Construction.}
To enable efficient retrieval, we first build a dual-view index for all users. For each user $v$, we construct a textual sketch $\psi_v$ by concatenating their inferred intents (Sec.~\ref{ssec:intent-polarity}) and recent item keyphrases (Sec.~\ref{ssec:item-semantic}). We store:
(1) A \textbf{Dense View} $\mathbf{m}_v = f_{\text{text}}(\psi_v)$, where $f_{\text{text}}$ is a frozen, pre-trained text encoder (e.g., Sentence-BERT). This captures semantic similarity.
(2) A \textbf{Sparse View} $\mathbf{s}_v$, representing the bag-of-words vector of the user's history, used to capture exact keyword matches.

\textbf{Hybrid Retrieval.}
For a target user $u$, we generate their query vector $\mathbf{m}_u$ using the same frozen encoder $f_{\text{text}}$. We then retrieve the top-$K$ neighbors by combining dense and sparse similarities:
\begin{equation}
\mathrm{sim}(u,v)=\lambda \frac{\mathrm{BM25}(\mathbf{s}_u, \mathbf{s}_v)}{Z_{\mathrm{bm}}} + (1-\lambda)\cos(\mathbf{m}_u, \mathbf{m}_v).
\label{eq:hybrid}
\end{equation}
This hybrid approach ensures that retrieved neighbors are similar both in broad semantic intent and in specific item preferences. Finally, to prevent information redundancy in the limited context window, we re-rank the retrieved neighbors using Maximal Marginal Relevance (MMR) before serializing their summaries into the final prompt.

\subsection{Reasoning-based Interest Matching}
\label{ssec:reasoning-score}
In the final stage, we consolidate all signals into a structured prompt that guides the LLM to perform evidence-grounded reasoning rather than simple pattern matching.

We construct the evidence set $\mathcal{E}(u,i)$ containing: (1) the user's explicit intent keywords and polarity; (2) the retrieved collaborative summaries from similar users; and (3) the candidate item's keyphrases. To aid the LLM, we also compute and append explicit \emph{coverage features}:
\begin{equation}
\mathrm{cov}^{\pm}(u,i)=\max_{w\in\mathcal{K}^{\pm}_u}\ \max_{k\in S_i}\ \cos(\mathbf{e}_w,\mathbf{e}_k).
\label{eq:cov}
\end{equation}
These features quantify the alignment between the user's top polarity keywords $\mathcal{K}^{\pm}_u$ and the item's attributes $S_i$.

The LLM is instructed to act as a judge. It analyzes $\mathcal{E}(u,i)$, verifies whether the item matches the user's inferred intent, and outputs a categorical verdict (No/Partial/Strong Match) supported by a concise rationale. We map the output token probabilities to a scalar score $\tilde{s}(u,i)$, which is calibrated via temperature scaling. This approach effectively decouples the reasoning process from pure memorization, resulting in robust performance even for long-tail items.

\section{EXPERIMENTS AND RESULTS}
\label{sec:pagestyle}

\begin{table*}[t]
  \centering
  \caption{Main comparison on \textit{ML-1M}, \textit{Games}, and \textit{Bundle}. Higher is better. Best results are in \textbf{bold}; second-best are marked with $^{\ast}$.}
  \label{tab:main}
  \begingroup
  \renewcommand{\arraystretch}{1.3}
  \setlength{\tabcolsep}{2.8pt}
  \scriptsize
  \scalebox{0.95}{%
  \begin{tabular}{llcc ccc ccc cccc c c c}
    \hline\hline
     &  & \multicolumn{2}{c}{Conventional Methods}
        & \multicolumn{3}{c}{Single-Intent Methods}
        & \multicolumn{3}{c}{Multi-Intent Methods}
        & \multicolumn{4}{c}{LLM Methods}
        & \multicolumn{1}{c}{Ours}
        & \multicolumn{1}{c}{Improve}
        & \multicolumn{1}{c}{$p$-value} \\
    \cline{3-4}\cline{5-7}\cline{8-10}\cline{11-14}\cline{15-15}\cline{16-16}\cline{17-17}
    Data & Metrics
        & MostPop & FPMC
        & NARM & STAMP & GCE-GNN
        & MCPRN & HIDE & Atten-Mixer
        & NIR & PO4ISR & RecMind & UniRec
        & \textbf{R\textsuperscript{3}-REC}
        &  &  \\
    \hline
    \multirow{3}{*}{\textit{ML-1M}}
      & HR@1   & 0.0004 & 0.1132 & 0.1692$^{\star}$ & 0.1584 & 0.1312 & 0.1434 & 0.1498 & 0.1490 & 0.0572 & 0.2000 & \textbf{0.2700}$^{*}$ & 0.2650 & \textbf{0.287} & +6.3\% & $4.3\mathrm{e}{-3}$ \\
      & HR@5   & 0.0070 & 0.3748 & 0.5230$^{\star}$ & 0.5078 & 0.4748 & 0.4788 & 0.4998 & 0.4932 & 0.2326 & 0.5510 & 0.5720 & \textbf{0.5910}$^{*}$ & \textbf{0.597} & +1.0\% & $5.9\mathrm{e}{-2}$ \\
      & HR@10  & 0.0120 & 0.5150 & 0.7050$^{\star}$ & 0.6920 & 0.6600 & 0.6640 & 0.6820 & 0.6760 & 0.3550 & 0.7240 & 0.7480 & \textbf{0.7620}$^{*}$ & \textbf{0.769} & +0.9\% & $4.6\mathrm{e}{-2}$ \\
    \hline
    \multirow{3}{*}{\textit{Games}}
      & HR@1   & --     & 0.0498 & 0.0572 & 0.0556 & 0.0692 & 0.0522 & 0.0696 & 0.0530 & 0.1168$^{\star}$ & 0.2588 & \textbf{0.3450}$^{*}$ & 0.3400 & \textbf{0.379} & +9.9\% & $6.4\mathrm{e}{-5}$ \\
      & HR@5   & --     & 0.2564 & 0.2574 & 0.2586 & 0.2744 & 0.2416 & 0.2694 & 0.2472 & 0.3406$^{\star}$ & 0.5866 & \textbf{0.6550}$^{*}$ & 0.6450 & \textbf{0.679} & +3.7\% & $3.5\mathrm{e}{-5}$ \\
      & HR@10  & --     & 0.4020 & 0.4050 & 0.4060 & 0.4260 & 0.3900 & 0.4180 & 0.3980 & 0.5100 & 0.7600 & \textbf{0.7920}$^{*}$ & 0.7800 & \textbf{0.812} & +2.5\% & $1.7\mathrm{e}{-5}$ \\
    \hline
    \multirow{3}{*}{\textit{Bundle}}
      & HR@1   & --     & 0.0398 & 0.0322 & 0.0365 & 0.0360 & 0.0360 & 0.0458 & 0.0525 & 0.0975$^{\star}$ & 0.1697 & \textbf{0.2550}$^{*}$ & 0.2450 & \textbf{0.281} & +10.2\% & $2.0\mathrm{e}{-5}$ \\
      & HR@5   & 0.0042 & 0.2475 & 0.2332 & 0.2352 & 0.2237 & 0.2352 & 0.2585 & 0.2644 & 0.2832$^{\star}$ & 0.4328 & \textbf{0.5150}$^{*}$ & 0.5050 & \textbf{0.548} & +6.4\% & $2.6\mathrm{e}{-4}$ \\
      & HR@10  & 0.0070 & 0.3820 & 0.3650 & 0.3680 & 0.3500 & 0.3680 & 0.3940 & 0.4020 & 0.4300 & 0.6120 & \textbf{0.6650}$^{*}$ & 0.6550 & \textbf{0.703} & +5.7\% & $1.8\mathrm{e}{-4}$ \\
    \hline\hline
     \vspace{-6 mm}
  \end{tabular}
  }
  \endgroup
\end{table*}

\subsection{Experiment settings}

We adopt a unified protocol: user histories are truncated to $H_{\max}{=}100$ and recommendations use top-$k$ scoring over a fixed 20-candidate pool (constructed per session by including the ground-truth next item and randomly sampling the remaining items, following PO4ISR).
The LLM backbone is GPT-3.5-Turbo with one deterministic query per user.
We report Hit-Rate@k (HR@1/5/10) and use an 8:1:1 train/validation/test split.

\textbf{Datasets.}
We use three real-world datasets spanning ratings and session logs: \emph{ML-1M}~\cite{Harper2015MovieLens} (movie ratings), \emph{Games}~\cite{Ni2019Justifying} (the Amazon video-games subset), and \emph{Bundle}~\cite{Sun2022RevisitBundle} (Amazon sessions from Electronics/Clothing/Food with crowdsourced intent annotations). For \emph{ML-1M} and \emph{Games}, interactions are chronologically ordered and segmented into day-level sessions; \emph{Bundle} provides sessions directly. We split sessions into train/validation/test with an 8:1:1 ratio.

\textbf{Baselines.}
We categorize prior art by modeling principle. 
\emph{Latent sequential encoders}—RNN/Transformer pipelines (GRU4Rec, SASRec, BERT4Rec) and graph variants (SR-GNN, GCE-GNN)— scale but hide semantics and struggle under sparsity~\cite{Hidasi2016GRU4Rec,Kang2018SASRec,Sun2019BERT4Rec,Wu2019SRGNN,Wang2020GCEGNN}, prompting recent diffusion-based methods to address multimodal feature denoising~\cite{lu2025dmmd4sr,cui2025diffusion,cui2025multi}. 
\emph{Intent/multiinterest models} (NARM, STAMP, MCPRN, HIDE, Atten-Mixer) disentangle factors to lift diversity, yet often fix intent cardinality or underuse textual cues~\cite{Li2017NARM,Liu2018STAMP,Wang2019MCPRN,Chai2022HIDE,Zhang2023AttenMixer}. 
\emph{LLM based recommenders} (NIR, PO4ISR, RecMind) add language priors via prompting or interactive generation to improve coherency~\cite{Hou2023NIR,Sun2024PO4ISR,Wang2024RecMind,xie2025chat,liu2025coherency}, while strong non-LLM regularization (UniRec) remains competitive~\cite{UniRec2024}. 
No strand jointly delivers fine-grained intent reasoning, robust text denoising, cold start resilience, and end-to-end explainability.

\subsection{Overall Performance Comparison}
As shown in Table~\ref{tab:main}, \textbf{R\textsuperscript{3}-REC} attains the best top-$K$ ranking quality on \textit{ML-1M}, \textit{Games}, and \textit{Bundle}. On \textit{ML-1M}, it surpasses the strongest LLM baselines (RecMind-25/UniRec-25) by {+}6.3\% HR@1, {+}1.0\% HR@5, and {+}0.9\% HR@10 ($p=4.3{\times}10^{-3}$, $5.9{\times}10^{-2}$, and $4.6{\times}10^{-2}$). The gains are larger on \textit{Games} ({+}9.9\% HR@1, {+}3.7\% HR@5, {+}2.5\% HR@10; $p\le 1.7{\times}10^{-5}$) and on \textit{Bundle} ({+}10.2\% HR@1, {+}6.4\% HR@5, {+}5.7\% HR@10; $p\le 2.6{\times}10^{-4}$). Conventional, single-intent, and multi-intent neural baselines lag markedly behind, indicating that reasoning-driven, multi-granular signals yield more precise top-$K$ ranking. Overall, improvements are statistically significant on \textit{Games} and \textit{Bundle} for all metrics, and on \textit{ML-1M} for HR@1 and HR@10 (HR@5 is marginal), demonstrating robust effectiveness of \textbf{R\textsuperscript{3}-REC} across domains.

\subsection{Ablation Studies}
\begin{figure}[H]
  \vspace{-2 mm} 
  \centering
  \includegraphics[width=0.8\linewidth]{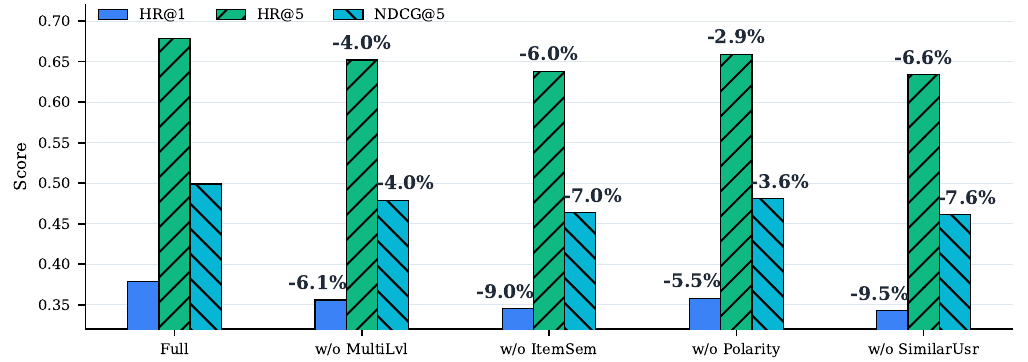}
  \vspace{-2 mm}
  \caption{Ablation on R\textsuperscript{3}-REC (HR@1 / HR@5 / NDCG@5). Lower bars on variants indicate performance drops relative to the full model.}
  \label{fig:ablation-bars}
  \vspace{-2 mm} 
\end{figure}

We ablate four components of \textbf{R\textsuperscript{3}-REC} on the \textit{Games} validation split (seed 0), removing one module at a time and visualizing HR@1/HR@5/NDCG@5 in Fig.~\ref{fig:ablation-bars}. Eliminating \emph{item–semantic extraction} yields the largest degradation, with HR@1 decreasing by about 9--10\% and HR@5/NDCG@5 by roughly 6--7\%, highlighting the role of content-aware cues for long-tail items. Removing \emph{similar-user enhancement} produces comparable declines (about 9--10\% on HR@1 and 6--8\% on HR@5/NDCG@5), evidencing the value of collaborative signals under sparsity. \emph{Multi-level intent modeling} remains material, reducing HR@1 by around 6\% and HR@5/NDCG@5 by about 4\%. \emph{Polarity mining} contributes complementary gains, with approximate 5--6\% and 3--4\% drops on HR@1 and HR@5/NDCG@5, respectively. The consistent, sub-additive declines indicate that these modules capture distinct yet synergistic signals for robust top-\emph{K} ranking.

\subsection{Hyperparameter Sensitivity }
\label{sec:rq3}

\begin{figure}[H]
  \centering
  \begin{subfigure}{0.8\linewidth}
    \centering
    \includegraphics[width=\linewidth]{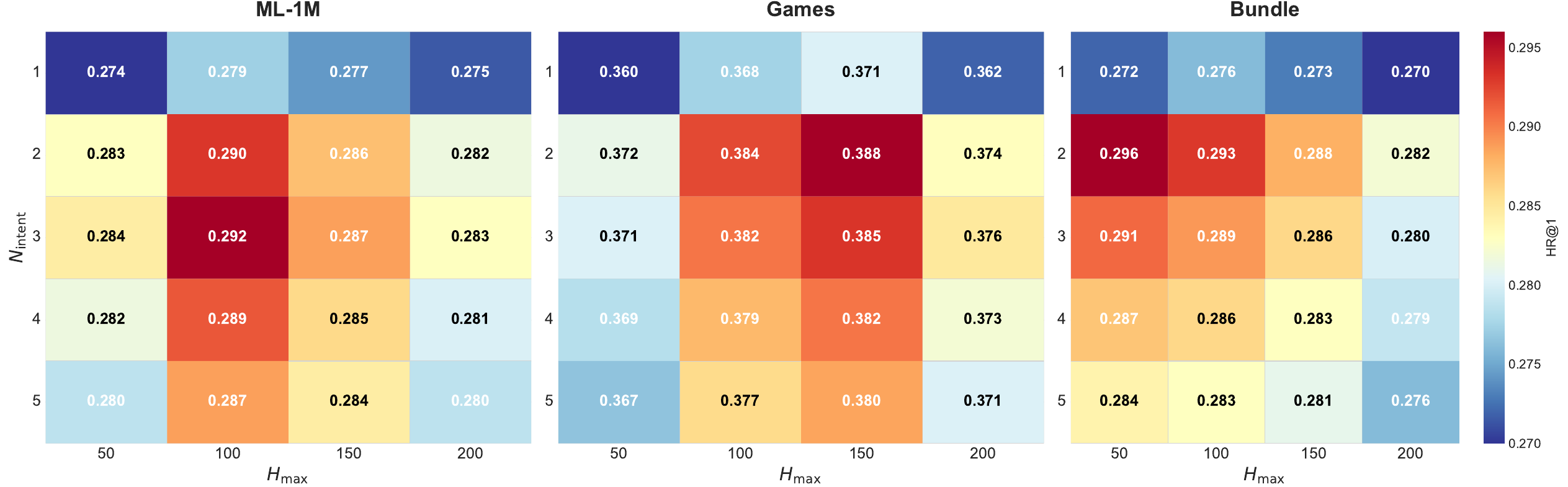}
    \caption{HR@1 vs.\ $N_{\text{intent}}$ and $H_{\max}$.}
    \label{fig:ablation-a}
  \end{subfigure}
  \vspace{-1mm}
  \begin{subfigure}{0.8\linewidth}
    \centering
    \includegraphics[width=\linewidth]{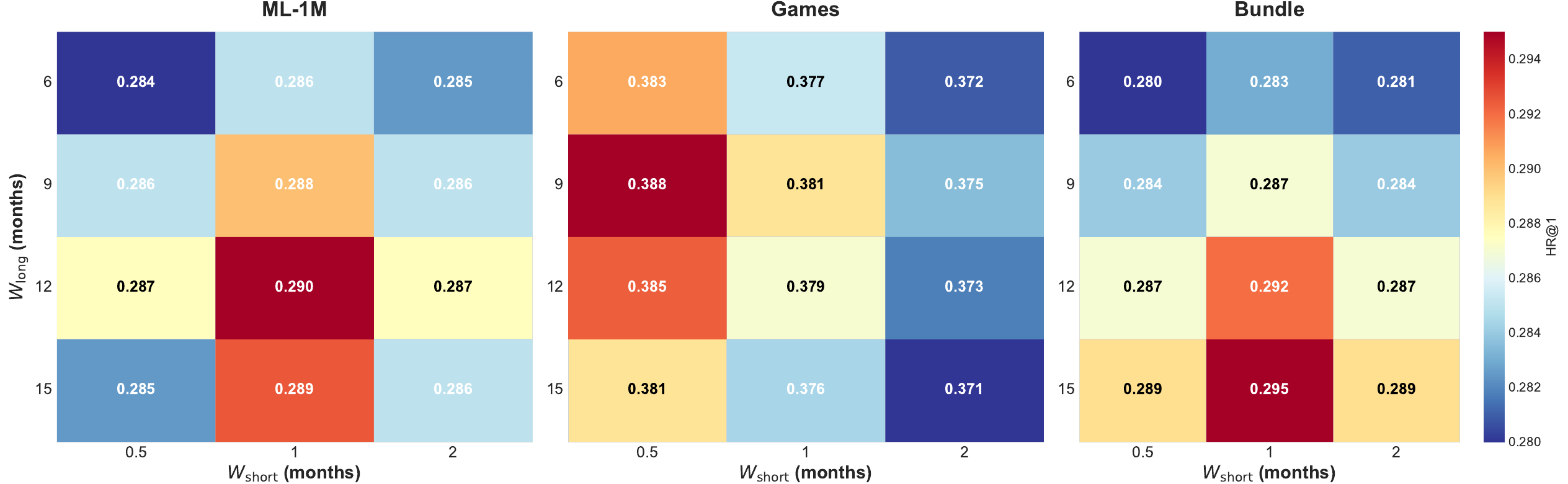}
    \caption{HR@1 vs.\ $W_{\text{long}}$ and $W_{\text{short}}$.}
    \label{fig:ablation-b}
  \end{subfigure}
  \vspace{-2mm}
  \caption{Hyperparameter sensitivity under two factorizations.}
  \label{fig:ablation}
  \vspace{-2mm}
\end{figure}

\textbf{User-intent count vs.\ history depth.}
We grid-search $N_{\text{intent}}\!\in\!\{1,2,3,4,5\}$ and $H_{\max}\!\in\!\{50,100,150,200\}$. 
Peaks are dataset-specific: \textit{ML-1M} at \textbf{($3,100$)}, \textit{Games} shifting deeper to \textbf{($2,150$)}, and \textit{Bundle} favoring short histories at \textbf{($2,50$)}. 
Using more than three intents introduces redundancy, and very long windows ($\ge\!200$) revive stale/conflicting evidence (Fig.~\ref{fig:ablation-a}).

\noindent\textbf{Long/short temporal windows.}
Sweeping $W_{\text{long}}\!\in\!\{6,9,12,15\}$ and $W_{\text{short}}\!\in\!\{0.5,1,2\}$, 
\textit{ML-1M} forms a ridge near \textbf{($12,1$)}, \textit{Games} peaks at \textbf{($9,0.5$)}, and \textit{Bundle} at \textbf{($15,1$)}. 
Departures change HR@1 by $\le\!1.8$\,pp, indicating robustness (Fig.~\ref{fig:ablation-b}).

\noindent\textbf{Keyword span \& neighbor threshold.} Using 5–8 keyphrases per item maximizes coverage without prompting overhead (raising the minimum from 3 to 5 adds $\approx\!1.2$\,pp HR@1; $>\!8$ yields no gains). Once $k_{\text{sim}}$ is tuned, varying the BM25 cut-off in $[0.25,0.45]$ changes HR@1 by $<\!0.2$\,pp.

\subsection{Efficiency and Cost Analysis}
\label{sec:efficiency}

We analyze the trade-off between accuracy, latency, and cost (Table~\ref{tab:efficiency_cost}).
\textbf{Latency \& ROI.} While \textsc{NIR} is fastest, it lacks precision. The agent-based \textsc{RecMind} suffers high latency ($\approx$920\,ms) due to the bottlenecked decoding of long Chain-of-Thought outputs.
In contrast, \textbf{\textsc{R\textsuperscript{3}-REC} achieves $\approx$580\,ms}. By shifting complexity to the input side (retrieval prefilling is parallelizable) and minimizing output to scoring tokens, we maintain sub-second responsiveness.
Although retrieval increases input tokens ($\approx$1.2k), \textsc{R\textsuperscript{3}-REC} delivers a massive accuracy leap (+9.9\% HR@1 vs.\ \textsc{RecMind}). This represents a superior return on investment (ROI) compared to baselines that expend latency budgets on slow reasoning generation.

\begin{table}[H]
  \centering
  \caption{Efficiency Analysis (20-item slate). \textit{Est. Cost}: Based on GPT-3.5 pricing. \textit{Latency}: Ranking phase only.}
  \label{tab:efficiency_cost}
  \scriptsize
  \renewcommand{\arraystretch}{1.0} 
  \setlength{\tabcolsep}{0pt}      

  \begin{tabular*}{\columnwidth}{@{}@{\extracolsep{\fill}}l c c c c c@{}}
    \toprule
    \multirow{2}{*}{Model} & \multicolumn{2}{c}{Avg. Tokens} & \multirow{2}{*}{Latency} & \multirow{2}{*}{Est. Cost} & \multirow{2}{*}{HR@1} \\
    \cmidrule(lr){2-3}
     & Input & Output & (ms) & (\$) & (Games) \\
    \midrule
    NIR \cite{Hou2023NIR}            & ~150 & ~10 & 250 & $<$1e-4 & 0.1168 \\
    PO4ISR \cite{Sun2024PO4ISR}      & ~350 & ~15 & 380 & 2e-4   & 0.2588 \\
    RecMind \cite{Wang2024RecMind}   & ~600 & ~80 & 920 & 4e-4   & 0.3450 \\
    \textbf{\textsc{R\textsuperscript{3}-REC}} & \textbf{1250} & \textbf{~10} & \textbf{580} & \textbf{6e-4} & \textbf{0.3790} \\
    \bottomrule
  \end{tabular*}

  \vspace{-4mm}
\end{table}

\section{CONCLUSION}
\label{sec:typestyle}

We introduced \textsc{R\textsuperscript{3}-REC}, a prompt-centric, reasoning-augmented recommender that unifies Multilevel User Intent Reasoning, Item Semantic Extraction, Long-Short Interest Polarity Mining, Similar User Collaborative Enhancement, and Reasoning-based Interest Matching and Scoring. Across ML-1M, Games, and Bundle, R\textsuperscript{3}-REC consistently improves top-K ranking with statistically significant gains, while preserving acceptable latency. Ablations confirm complementary benefits: removing Item Semantic Extraction or Similar User Collaborative Enhancement yields the largest drops, whereas Multilevel User Intent Reasoning and Long-Short Interest Polarity Mining supply stable, interpretable cues. Sensitivity analyses further indicate robust behavior under decoding temperature, intent cardinality, history depth, and temporal windows. Overall, R\textsuperscript{3}-REC demonstrates that explicit, retrieval-augmented reasoning over multi-granular signals is both effective and practical. In the future, we will use a lighter non-LLM scorer with the same engineered signals.

\vfill\pagebreak

\bibliographystyle{IEEEbib}
\bibliography{strings,refs_cr}

\end{document}